\documentclass[prb,aps,amssymb,showpacs,twocolumn,amsmath,floatfix]{revtex4}
\usepackage{tabularx}
\usepackage{bm}
\usepackage{euscript}
\usepackage{epsfig,psfrag,subfigure}
\usepackage{graphicx}
\usepackage{color}
\usepackage{amsfonts}
\usepackage{exscale}
\usepackage{amsbsy}
\usepackage{hyperref}
\input{epsf}

\def\be{\begin{equation}}
\def\ee{\end{equation}}
\def\ba{\begin{eqnarray}}
\def\ea{\end{eqnarray}}

\def\SRO{Sr$_3$Ru$_2$O$_7$}

\begin{document}

\title{Microscopic theory of the nematic phase in \SRO}
\author{S. Raghu$^1$, A. Paramekanti$^2$, E.-A. Kim $^3$, R. A. Borzi$^{4,5}$, S. A. Grigera$^{4,6}$, A. P. Mackenzie$^4$, and S. A. Kivelson$^1$}
\affiliation{$^1$Department of Physics, Stanford University, Stanford, CA 94305}
\affiliation{$^2$Department of Physics, University of Toronto, Toronto, Ontario M5S 1A7,
Canada}
\affiliation{$^3$Department of Physics, Cornell University, Ithaca, NY 14853}
\affiliation{$^4$Scottish Universities Physics Alliance, School of Physics and
Astronomy, University of St. Andrews, North Haugh, St. Andrews KY16
9SS, UK}
\affiliation{$^5$Instituto de Investigaciones Fisicoqu\'imicas Te\'oricas
y Aplicadas, and Departamento de F\'isica (UNLP), IFLP (CONICET), 1900 La Plata,
Argentina.}
\affiliation{$^6$Instituto de F\'{\i}sica de L\'{\i}quidos y Sistemas
 Biol\'ogicos, UNLP, La Plata 1900, Argentina}
\date{\today}

\begin{abstract}
In an externally applied
magnetic field, ultra-pure crystals of the  
bilayer 
 compound ${\rm Sr}_3{\rm Ru}_2{\rm O}_7$ undergo a metamagnetic
transition below a critical temperature, $T^*$, which 
varies as a function of the angle between the magnetic field $H$ and the Ru-O planes.  Moreover,
$T^*$
approaches zero when $H$ is perpendicular to the planes. This putative ``metamagnetic quantum critical
point", however, is preempted by a nematic fluid phase with order one resistive
anisotropy in the {\it ab} plane.  
In a ``realistic'' bilayer model   
with moderate strength local
Coulomb interactions,  
the existence of a sharp divergence of the
electronic density of states near a van Hove singularity of the quasi-one-dimensional bands, and 
the presence of spin-orbit coupling
results in a mean-field
phase diagram which accounts for many of 
these experimentally observed
phenomena.  Although the spin-orbit coupling is not overly strong, it destroys
the otherwise near perfect Fermi surface nesting 
and hence suppresses spin-density-wave (SDW) ordering.

\end{abstract}


\maketitle

The Ruddlesden-Popper series ${\rm Sr}_{n+1}{\rm Ru}_n{\rm O}_{3n +1}$ is a family 
of ruthenate materials which exhibit a wide variety of electronic properties ranging from 
unconventional superconductivity ($n=1$) to itinerant electron ferromagnetism ($n \ge 3 $), and 
have been the focus of intense research for over a decade\cite{Mackenzie2003, Bergemann2003, Ovchinnikov2003}.  Most materials in this class are 
``bad metals\cite{Emery1994}" (the resistivity $\rho(T) \propto T$ and in excess of the Ioffe-Regel limit) at room temperature and above; yet, they obey Fermi-liquid
 theory at low temperatures (typically $T < 50 K$).  The low energy electronic 
 properties of these materials are determined mainly by the electrons in the ${\rm Ru}$ $t_{2g}$ 
 subspace consisting of the nearly degenerate $d_{xz},d_{yz},d_{xy}$ 
 orbitals.  Therefore, in addition to the spin and charge degrees of freedom, the orbital 
 degrees of freedom play an important role in determining the properties of these systems.
  Here, we focus on  the bilayer ($n=2$) compound 
 ${\rm Sr}_3{\rm Ru}_2{\rm O}_7$ which
 is neither a superconductor nor a ferromagnet.

\begin{figure}[t]
\includegraphics[width=5.6cm]{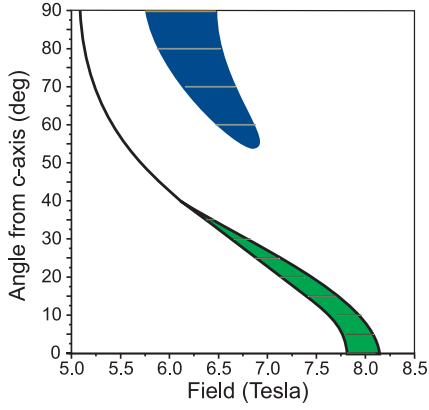}
\caption{(Color online) The experimentally determined low temperature phase diagram of \SRO in the field - angle plane, based on resistivity and magnetic susceptibility measurements on shape-unbiased octagonal crystals at 100 mK.  For the resistivity measurements, both the in-plane field component and the current are along either the crystallographic ${\it a}$ or ${\it b}$ axes.  The shaded regions are those in which a resistive anisotropy is observed.  The solid black lines represent first-order phase transitions as determined by a sharp dissipative peak in a measurement of the imaginary component of the a.c. susceptibility [see discussion in Grigera {\it et al.} \cite{Grigera2003}].  We observed no dissipative peak at the boundaries of the blue region, but cannot rule out its existence beyond our resolution. }
\label{expt}
\end{figure}

\SRO  is a tetragonal material consisting of 
RuO$_2$ planes forming
 bilayers which are stacked and weakly coupled to one another. 
In crystals of high purity 
 and structural perfection, a metamagnetic transition \cite{Perry2001}, i.e. a sudden and sharp rise in the 
 magnetization with a modest increase in the applied field, is observed.  
 While this transition is first-order, 
 the transition line terminates at a critical point $(H^*,T^*)$, 
 where it becomes continuous (in analogy with a liquid-vapor 
 transition in the pressure-temperature plane).  However, 
   the critical field and temperature $H^*$ and $T^*$,
 depend on the angle, $\theta$, between the magnetic field and the crystalline c-axis, perpendicular to the  RuO$_2$ bilayers:
$H^*$ decreases from $\sim 7.8 T$ to $\sim 5.1 T$ , as $\theta$ increases from 0 to $90 \textordmasculine$, while $T^*$ drops from $T^*( \theta = 90\textordmasculine)=1.25K$ to $T^* \sim 0$ as $\theta \to 0$, that is when $H$ is perpendicular to the RuO$_2$ planes \cite{Grigera2003}. 
 Thus, it was 
 proposed that this material exhibits a new type of quantum critical phenomena 
 associated with the termination point of the first order line of metamagnetic transitions \cite{Grigera2001}.  
 However, experiments involving ultra-pure, single crystal samples (with a residual resistivity less 
 than $1 \mu \Omega{\rm cm}$) 
 have shown that instead of such a ``metamagnetic quantum critical point" \cite{Millis2002}, 
 there is a bifurcation of the metamagnetic phase boundary, which leads to {\it two} first order
 metamagnetic transitions at closely spaced field values $H_{c1} \approx 7.8T$ and $H_{c2}
 \approx 8.1T$ \cite{Perry2004}.
 At
 intermediate fields, $H_{c1} < H < H_{c2}$, an electron nematic phase
 \cite{Kivelson1998,Oganesyan2001} appears, which 
 spontaneously breaks the discrete square lattice rotational symmetry from C$_4$ to C$_2$ as
 inferred from the observation of resistive anisotropy in the
 {\it ab} plane \cite{Borzi2007}.
The nematic phase occurs in a narrow range of fields, and for a range of angles $ 0 \le \theta \le 40 \textordmasculine$ (the green region in Fig. \ref{expt}).   Resistive anisotropy is also 
found in the blue region in Fig. \ref{expt}, for $ 55 \textordmasculine \le \theta \le 90 \textordmasculine $, 
but it is not known whether this reflects the existence of a new phase.

In the present paper, we study the microscopic origins of the weak metamagnetism and the accompanying nematicity.  
A possible microscopic route to understanding metamagnetism in this material was proposed by Binz and
Sigrist \cite{Binz2004} in a model  of a 
two dimensional band on a square lattice whose Fermi surface lies close to a van Hove (vH)
singularity.  
Incorporating weak local Coulomb repulsion between the electrons, they showed
that when the magnetic field tunes the Fermi surface of one spin species
close enough to the vH singularity, there is a jump in magnetization. 
Grigera et al \cite{Grigera2004}  were the first to propose that the existence of the vH singularities might be a driver for nematicity accompanying metamagnetism in a spin-dependent version of the Pomeranchuk distortion\cite{Wu2007} previously studied in the two-dimensional Hubbard and t-J models 
\cite{Yamase2000, Halboth2000}.  
A critical insight into the problem of nematicity accompanying metamagnetism was proposed in a paper by H.-Y. Kee and Y-.B. Kim \cite{Kee2005}, who 
 showed that additional interactions between the electrons can
lead to new instabilities which split the metamagnetic transition, leading
to an intermediate phase. Specifically, they considered a model with weak quadrupolar
interactions and showed that a mean field treatment
naturally leads to a sequence of two transitions as a function of increasing $H$. First, at a critical field $H_{c1}$, the spin-up Fermi surface reconnects  across the vH point at one edge of the Brillouin zone, leading to a metamagnetic jump  accompanied by the
spontaneous breakdown of $C_4$ symmetry (depending on which vH point reconnects).  
Then, at a   higher field $H_{c2}$, the spin-up Fermi surface reconnects across the second vH point, leading to a second metamagnetic jump and a restoration of   $C_4$ symmetry.

The model originally considered by Kee and Kim\cite{Kee2005}  is a single  band model, with strongly $\vec k$ dependent effective interactions, engineered to promote nematicity.  
While it ties metamagnetism and nematicity in an ingenious way, it still 
leaves open the issue of the possible microscopic origins of such phenomena.  In particular, 
it does not address the issue of what features of the material's electronic structure are most 
important in accounting for its phase diagram.  

\begin{figure}
\includegraphics[width=8.cm]{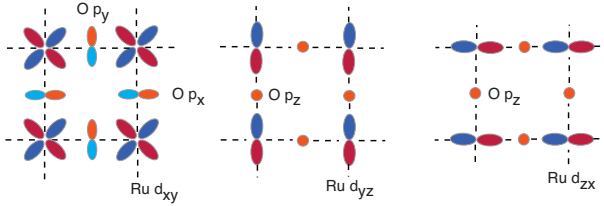}
\caption{(Color online) The Ru $d_{xy},d_{yz},d_{xz}$ orbitals in a single layer of \SRO in the {\it ab} plane.  
$\pi$-overlaps between two like orbitals along nearest-neighbor bonds are mediated by the 
intervening oxygen p-orbitals.  For two identical $d_{ij}$ orbitals ($i \ne j$, and $ i,j=x,y,z$), the hopping 
along the $\hat{i}$ direction is strong and is mediated by the oxygen $p_j$ orbital and {\it vice-versa}.  $\delta$-overlaps between two identical d-orbitals along a nearest 
neighbor bond do not make use of the oxygens, and are therefore much weaker in comparison.  
Furthermore, all nearest-neighbor hopping between two distinct d-orbitals vanish by symmetry.}
   \label{orbitals}
\end{figure}

Here, we address this issue by   
considering a more realistic model of the 
electronic structure of \SRO.  Since both band-structure calculations and angle-resolved photoemission
 (ARPES) experiments show that there are at least three bands at the Fermi energy per Ru atom (6 per bilayer),  we consider a model (Section \ref{model}) with three Wannier functions, corresponding loosely to the  Ru d$_{xz}$, d$_{yz}$, and d$_{xy}$ orbitals (Fig. \ref{orbitals}).  The former two give rise (as we discuss in Section 
  \ref{meanfield}) to crisscrossing quasi 1D bands, while the latter gives rise to quasi-2D bands.  
  Each of these bands is, moreover, split in two by the inter-bilayer hopping, which is substantial 
  in the case of the quasi-1D bands.  
  The multi-orbital band-structure also implies the existence of an on-site spin-orbit coupling of moderate strength.  We study the ordered phases produced by physically reasonable local (on site) Coulomb interactions using unrestricted Hartree-Fock wavefunctions (See Fig. \ref{phasediagram}.)  
  
In the multiband context,  the nematic phase corresponds to a particular {\it orbital-ordered}
   broken symmetry configuration.  
  Moreover, we will show below that transitions into such a 
  nematic phase are naturally accompanied by metamagnetic transitions.  
  Because the 1D bands are closer to the vH points, and because of the stronger divergence of the
  density of states (DOS), $\nu$,  at the vH point in 1D, ($\nu \sim 1/\sqrt{E}$), we find that both 
  nematicity and metamagnetism order are primarily driven by a collective reordering of these bands.  
 In addition, a number of other qualitative features of the experimentally observed phase diagram occur naturally and generically from the mean-field solution of the present model:
   
 1)  The interval in which the nematic phase occurs can be tuned to be relatively small, $(H_{c2} - H_{c1})/H_{c2} \ll 1$.
 (See Fig. \ref{fig2}.)
 
2)  There is 
an asymmetry to the problem, apparent   in Fig. \ref{fig2} and in the experimental data, which results in a monotonic decrease of the nematic order as $H$ rises from $H_{c1}$ to 
   $H_{c2}$, resulting in a smaller change in the nematicity at $H_{c2}$ than at $H_{c1}$.  This 
  feature arises in our model due to the underlying asymmetry in the electronic density of states near the vH singularity of the quasi-1D bands.  
   
 3) Our model naturally accounts for why metamagnetism and nematic phases do not occur in the monolayer ruthenate 
 ${\rm Sr}_2{\rm Ru O}_4$.    The strong bilayer splitting in \SRO places the Fermi level 
 in a region where the density of states of the 1D bands has pronounced positive curvature, satisfying 
 the requirement of the Landau theory of a weakly first order metamagnetic transition \cite{Wohlfarth1962}.  
   
   4)  Even a moderate spin orbit coupling, $\lambda_{s.o.} \sim 0.2 t$, consistent with band structure estimates
   \cite{Haverkort2008, Liu2008}, produces an order 1 decrease of the critical fields as $\theta$ varies from $0$ to $90 \textordmasculine$.  (See Fig. \ref{angle}.)
   
   5)  The near-perfect nesting of the quasi-1D bands accounts for the most prominent peaks in the low energy magnetic structure factor observed in neutron scattering experiments at $H=0$\cite{Capogna2003, Ramos2008}.  However, the spin-orbit coupling is remarkably efficient at spoiling this nesting, thus plausibly explaining why spin density wave (SDW) order does not actually materialize (at least for $H=0$)\cite{Capogna2003, Ramos2008}.

Clearly, there are many aspects of the physics that are more subtle, and cannot be addressed, even qualitatively, at mean field level.  There is, after all, considerable evidence of the effects of strong quantum fluctuations associated with the narrowly preempted metamagnetic quantum critical point.  We will return to these shortcomings at the end of the paper.

\section{The Model}
\label{model}

\begin{figure}
\includegraphics[width=10.cm]{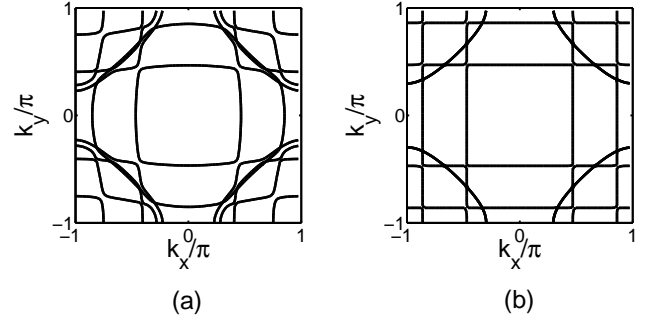}
\caption{ (a) A ``realistic" tight-binding Fermi surface of \SRO taking into account the Kinetic energy 
terms in $H_1$, $H_2$ as well as $H_{s.o.}$.  Here, $t'= 0.1 t$ and $\lambda_{s.o.} = 0.2 t$.  (b) Fermi-surface derived from an idealized model taking into 
account only the single particle terms in $H_1$.  In the nematic phase, one of the ``spin-up" bonding 
bands crosses the vH point at either $(\pi,0)$, or $(0, \pi)$ depending on the sign of the nematic order 
parameter (to be defined below).  Quotes are placed around ``spin-up" because in the presence of spin-orbit coupling, the magnetic field 
lifts the degeneracy of Kramers' doublets, and in this case, it is the appropriate pseudo-spin band which 
crosses the vH point.      }
   \label{fs}
\end{figure}

We have studied a simple tight-binding model of a single RuO$_4$ bilayer, with terms 
organized according to a hierarchy of scales:
\be
\label{h}
H=H_1 +H_{s.o.} +H_2+ \ldots
\ee
where $H_1$ contains the largest terms, which involve the most direct $\pi$-overlaps between Ru d-orbitals on
 nearest-neighbor sites: the largest hoppings between two neighboring identical $d_{\alpha\alpha^\prime}$ orbitals ($\alpha, \alpha^\prime=x,y,z$, $\alpha\neq \alpha^\prime$) 
 are along the crystalline $\hat{\alpha}$ and $ \hat{\alpha}^\prime$ directions.
 These hoppings in turn make use of the intervening oxygen p-orbitals.  
 The single bilayer approximation is a good one because transport measurements confirm the existence of highly two-dimensional transport, and therefore weak bilayer-bilayer coupling.  
 $H_1$ also includes the onsite Coulomb repulsion terms between two electrons on the 
 same orbital (U), as well as between two electrons in different orbitals (V).    
  $H_{s.o.}$ captures the effect of onsite 
 spin-orbit coupling.
 $H_2$ represents the kinetic energy terms due to weaker 
 $\delta$-overlaps between the orbitals (e.g. along the $\hat{z}$ 
 direction for the $d_{xy}$ orbitals).  Still smaller terms, some of which we will mention below, are represented by the ellipsis.

 Let
$d^\dagger_{\alpha, \sigma,  {\vec R}, \lambda}$ create an electron with spin polarization $\sigma$ 
 at horizontal position $\vec{R}$, in layer $\lambda = \pm 1$, and in orbital $\alpha=x,y,z$ 
 corresponding to the d$_{xz}$,  
 d$_{yz}$, or  d$_{xy}$ orbital respectively.  
In order to emphasize the underlying symmetries of the 
 Hamiltonian, we define
  a spinor field 
  \begin{equation}
  \Psi^{\dagger}_{\alpha \sigma}(\vec{R}, \lambda)= d^{\dagger}_{\alpha , \sigma , {\vec R} , \lambda} \ , \  \alpha = x,y,z .
  \end{equation}
  In terms of these, and  $n_{\lambda,\alpha,\vec  R}=\sum_{\sigma}d^{\dagger}_{\alpha , \sigma , {\vec R} , \lambda} d_{\alpha , \sigma , {\vec R} , \lambda} $,
\ba
H_1 &=& -t \sum_{\lambda, \vec{R} }  \Big[  \Psi^{\dagger}({\vec R}, \lambda)  \left( \hat{T}^x + \hat{T}^z \right) \Psi({\vec R}+ \hat{x}, \lambda) 
\nonumber \\
&&+ \Psi^{\dagger} ({\vec R}, \lambda)  \left( \hat{T}^y + \hat{T}^z \right) \Psi({\vec R}+ \hat{y}, \lambda)
\nonumber \\
 && + \frac{1}{2} \Psi^{\dagger}({\vec R}, - \lambda) \left( \hat{T}^x + \hat{T}^y \right)   \Psi({\vec R}, \lambda) + {\rm H.C.}  \Big]\nonumber \\
 &&+ \frac U 2 \sum_{\lambda,\alpha, {\vec R}}n_{\lambda,\alpha,\vec  R}^2 + \frac V 2 \sum_{\lambda,\alpha\neq \alpha^\prime {\vec R}}n_{\lambda,\alpha,\vec  R}n_{\lambda,\alpha^\prime,\vec  R} \nonumber \\
 &&-  \vec{H} \cdot \sum_{{\vec R}, \lambda} \Psi^{\dagger}({\vec R}, \lambda) \left( \vec{L} +  \vec{S} \right)  \Psi({\vec R}, \lambda) 
\ea
The contraction over the spinor subscript indices is  implied and we have defined the following matrices:
\ba
\hat{T}^i_{\alpha \beta; \sigma \sigma '} &=& \delta_{\alpha \beta} \delta^i_{\alpha} \delta_{\sigma \sigma'} \nonumber \\
{L}^i_{\alpha \beta; \sigma \sigma'} &=& \ell^i_{\alpha \beta} \delta_{\sigma \sigma'} \nonumber \\
{S}^i_{\alpha \beta; \sigma \sigma'} &=& \delta_{\alpha \beta} \tau^i_{\sigma \sigma'}
\ea
where 
$\vec{\ell}$ are the orbital angular momenta projected onto the $t_{2g}$ states (an explicit form of these are given in Section \ref{soc} ), 
and $\vec{\tau}$ 
are the Pauli matrices.  
The final term above is the Zeeman coupling to an external field.  We note that in the $t_{2g}$ subspace, the angular momentum is 
only partially quenched, since it is possible, for example, to form linear combinations of orbitals  $d_{xz} \pm i d_{yz}$ which are 
eigenstates of $\ell^z$.   Therefore, the external magnetic field will couple both to ${\vec L} $ and ${\vec S}$.

The on-site spin-orbit coupling Hamiltonian is 
\begin{equation}
\label{hso}
H_{s.o.} =  \lambda_{s.o.} \sum_{{\vec R}, \lambda} \Psi^{\dagger} ({\vec R}, \lambda) \left( \vec{L}  \cdot \vec{ S} \right) \Psi({\vec R}, \lambda) 
\end{equation}
and the smaller nearest-neighbor couplings are
\ba
H_2 &=&  -t' \sum_{{\vec R}, \lambda} \Big[\Psi^{\dagger} ( {\vec R} , \lambda) \hat{T}^x \Psi ({\vec R} + \hat{y}, \lambda) \nonumber \\
&& +\Psi^{\dagger} ( {\vec R} , \lambda) \hat{T}^y \Psi({\vec R} + \hat{x},\lambda)
\nonumber \\
&&+\frac{1}{2} \Psi^{\dagger}( {\vec R} , \lambda) \hat{T}^z \Psi ({\vec R}, -\lambda)  + {\rm H.C.} \Big ] .
\ea
So far, in writing the above terms, we have assumed locally perfect octahedral symmetry with the result, for example, that the hopping matrix element, $t$, between $d_{xz}$ orbitals on sites separated by one lattice constant in the x-direction, or on equivalent sites in neighboring planes of a bilayer are equal to each other.  
The actual material deviates slightly  from this ideal symmetry \cite{Shaked2000};
small terms that break this symmetry, as well as 
 further-range hopping terms and 
   interactions, are all represented schematically by the ellipsis 
  in Eq. \ref{h}, and will not be considered explicitly here.

 The Fermi surface of our model taking into account the kinetic terms of $H_{1}, H_{2}$ 
 and also the spin-orbit term $H_{s.o.}$ is shown in Fig. \ref{fs}a.  
  It is useful to contemplate its relation to the simpler
  Fermi surface (shown in 
 Fig. \ref{fs}b )  obtained by 
  setting the couplings in
  $H_2$, and $H_{s.o.}$ to 0.  
 In this simpler model, hoppings only along the 
  strongest bonds of each orbital are taken into account, and consequently, 
 there are perfect one-dimensional bands which are split by the 
 bilayer hopping and form the straight patches of the Fermi surface.  These Fermi sheets are 
 purely $d_{xz}, d_{yz}$ in orbital character.  
 Moreover, the
 2D bands  which come from the $d_{xy}$ orbital are degenerate at this level of approximation.  
 Were we to include the effects of the smaller hopping terms in $H_2$, a small splitting of the 2D bands, and a slight warping of the 1D bands would result, but there still would not  be any mixing between bands, and hence there would be multiple points at which two pieces of the Fermi surface 
 would cross one another.  When spin-orbit coupling is 
  included, 
 these degeneracies are lifted, as is evident from Fig. \ref{fs}a.  Therefore, $H_{s.o.} $ has the important qualitative effect of 
 changing the Fermi surface topology, even if its  
 magnitude is small.   We will study the consequences of 
 including $H_{s.o.}$ in Section \ref{soc}.  
 
  The d$_{xy}$ 
 bands play little active role in the physics discussed below, so we will for the most part ignore 
 these bands altogether, treating them instead simply as a particle reservoir which allows us to 
 perform calculations involving the remaining bands at constant chemical potential rather than 
 at constant density.

\section{Mean-field theory}
\label{meanfield}

\begin{figure}
\includegraphics[width=10.cm]{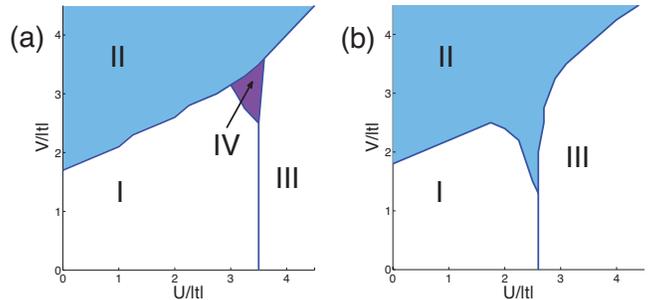}
\caption{ (Color online) Mean-field $T=0$ phase diagram in the $U- V$ plane in the absence 
of spin-orbit coupling for  $\mu = 0.81$, at zero 
applied field (a), and in finite field (b), taken here to be $|\vec H|=0.04t$.  
In zero field (a), there are four distinct phases: a paramagnetic region (I), a nematic region (shown in blue) (II), 
a region in which the ferromagnet and nematic spin nematic phases are degenerate, but do not 
coexist (III), and a region in which the nematic, ferromagnet, and nematic spin nematic phases coexist 
(purple region) (IV).  All phase boundaries correspond to first order transitions.  In a finite field (b), all regions acquire a non-zero magnetic moment.  Regions I and III do not 
have nematic order, whereas region II contains nematic order.  In this case, phase boundaries 
correspond to metamagnetic transitions.  For $U \approx V$ chosen such that at zero field, the parameters lie in region I in (a) above, a field sweep moves the phase boundaries closer to the origin so that the system traverses regions I $\rightarrow$ II $\rightarrow$ III  in (b) as the field increases.  In this way, the system exhibits a nematic phase precisely between two metamagnetic transitions.      }
   \label{phasediagram}
\end{figure} 

We first consider an idealized model of a bilayer consisting of 2 orbitals $(d_{xz}, d_{yz})$ in each layer.  
For the moment, we neglect the curvature of the Fermi surface and any mixing between the two orbitals.  
In this 
limit, the orbitals form perfect one-dimensional bands that are bilayer split due to the hopping along 
the c-axis.   
\begin{figure}
\includegraphics[width=8.cm]{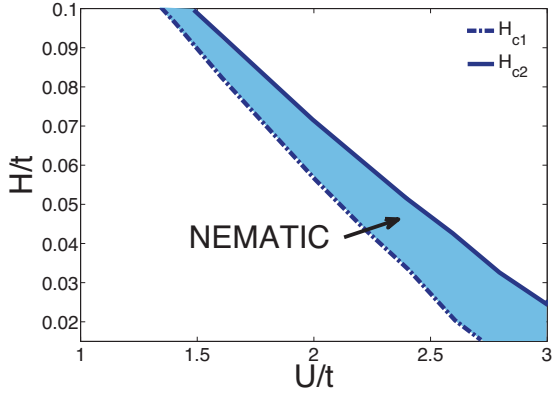}
\caption{ (Color online) Phase diagram at zero temperature in the  $U- H$ plane with $U=V$ in the absence 
of spin-orbit coupling for  $\mu = 0.81$.  For each $U=V$, as the field is increased, two metamagnetic transitions occur 
and are denoted $H_{c1}$ (dashed line) and $H_{c2}$ (solid line).  For $H_{c1} < H < H_{c2}$ a 
nematic phase (blue region) occurs, corresponds to region II in Fig. \ref{phasediagram}b,  and collapses onto region IV in Fig. \ref{phasediagram}a in the zero field limit.  
Although we have chosen $U=V$ in this figure, we emphasize that a phase diagram with the same 
topology occurs if we set $U= \alpha V$, for a range of $\alpha$, so long as the line $U= \alpha V$ crosses each of the three regions shown in Fig. \ref{phasediagram}b.    }
   \label{UvsH}
\end{figure} 
In  the 
presence of these multiple orbital degrees of freedom, it is natural to consider spin and orbital 
ordered broken symmetry states, and so we define the following sets of 
collective variables:
\begin{eqnarray}
\label{op}
  N_{\lambda} &=& \sum_{ \vec{R}} \langle \Psi^{\dagger}({\vec R} , \lambda) \left( \hat{T}^x + \hat{T}^y \right) \Psi({\vec R} , \lambda) \rangle \nonumber \\
  \vec{ M}_{\lambda} &=& \sum_{ \vec{R}} \langle \Psi^{\dagger}({\vec R} , \lambda) \left[ \vec{S} \left( \hat{T}^x + \hat{T}^y \right) \right] \Psi({\vec R} , \lambda) \rangle \nonumber \\
  N^o_{\lambda} &=& \sum_{ \vec{R}} \langle \Psi^{\dagger}({\vec R} , \lambda) \left( \hat{T}^x - \hat{T}^y \right) \Psi({\vec R} , \lambda) \rangle \nonumber \\
\vec{ N}^s_{\lambda} &=& \sum_{ \vec{R}} \langle \Psi^{\dagger}({\vec R} , \lambda) \left[ \vec{S} \left( \hat{T}^x - \hat{T}^y \right) \right] \Psi({\vec R} , \lambda) \rangle 
  \end{eqnarray}
These represent, respectively, the total electron density of the one-dimensional bands, $N$, their overall magnetization, $\vec{M}$ , 
the nematic order parameter,  $N^o$,  that represents the difference in charge density in the 
two bands, and lastly, we also define $\vec{N}^s$, which represents the difference in moment in each 
band.  We call this latter quantity the order parameter for the ``nematic spin nematic" phase \cite{Kivelson2003}.  
Whereas  $\vec{M}$ breaks time-reversal $(T)$ and $SU(2)$ symmetry,  and $N^o$ breaks the lattice 
$C_4$ rotation symmetry $(R)$, the nematic-spin-nematic order breaks $SU(2)$,  $(T)$ and $(R)$;  
however, the product $(TR)$ is not broken in this phase.  In the presence of an 
externally applied magnetic field, there is no distinction between the nematic and the nematic-spin-nematic phases.   
\begin{figure}
\includegraphics[width=10.cm]{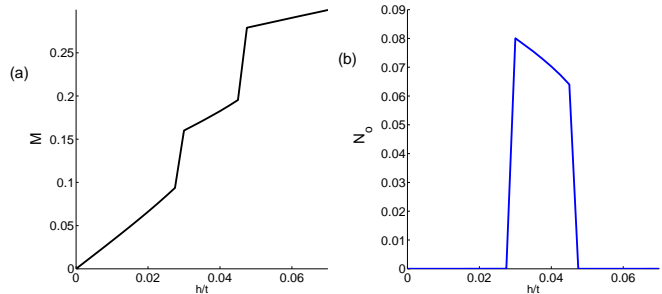}
\caption{ (Color online) (a) Magnetization and (b) nematicity as a function of applied field for $U=V=2.5t$, $\mu = 0.81t$
in the absence of spin-orbit coupling.  
The nematic phase is found to lie precisely between the 
two metamagnetic transitions.   }
   \label{fig2}
\end{figure} 

The mean-field phase diagram of our model is presented in Figure \ref{phasediagram}.  The 
details of the calculations are presented in the appendix.  The zero-field 
phase diagram (a) consists of 4 phases:  I: a paramagnetic phase, II: a nematic phase, III: a phase in 
which the ferromagnet and nematic-spin-nematic are degenerate but do not coexist, and IV: a 
coexistence phase of nematic, nematic-spin-nematic, and ferromagnetism.  All phase-boundaries 
shown here are first-order transitions.  
When the magnetic field is non-zero (b), the ferromagnetic phase has a lower free energy 
than the nematic-spin-nematic phase and this degeneracy is lifted.  

In a non-zero field, all phase boundaries 
become metamagnetic transitions.  Two qualitatively different features arise in this case.  First, there 
are only two types of phases: either magnetic order is present alone (I, III), or magnetic order coexists 
with nematic order (II).  In (b), the only difference between regions I and III is the size of the moment 
(larger in region III).  Note in particular, that with increasing magnetic field, phase III moves in to smaller 
values of U and V.  Thus, without requiring any fine-tuning, it is possible for the system studied here to 
undergo a sequence of metamagnetic transitions, starting with a paramagnetic phase, ending with a 
ferromagnetic phase, with nematic order sandwiched in between, as shown in Fig. \ref{UvsH}.   The requirements for such a 
phenomenon to occur in our model is that $U \approx V$.  For $U \gg V$, for instance, 
the effect of applying a magnetic field is to induce ferromagnetism alone.  By contrast, when 
$U \approx V$, this ferromagnetic phase is preempted at low fields by a nematic phase.  
Figure \ref{fig2} shows the magnetization and nematic order as a function of field for  a particular 
value of $U=V=2.5t$.  Here it is clearly seen that as the field sweeps across the first metamagnetic 
transition, nematic order develops; when the field is increased further, a second metamagnetic 
transition occurs which destroys the nematic phase.  

\begin{figure}
\includegraphics[width=8.cm]{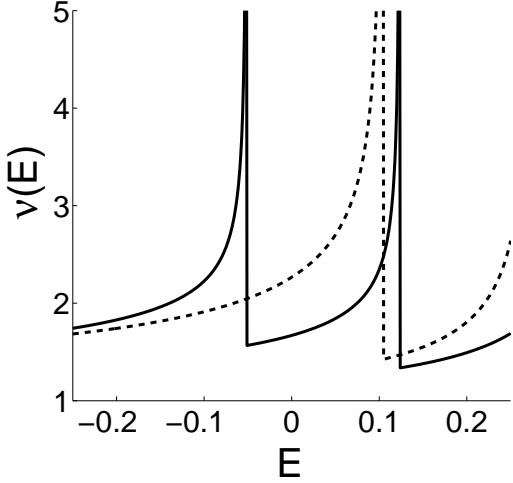}
\caption{ Single quasiparticle density of states (arbitrary units) in the nematic phase, $ H =0.04t, U=V=2.5t, \mu = 0.81t$.  
The Fermi energy is at $E=0$.  The dashed line corresponds to the quasiparticle density of states at 
  $H=0.02t$, there the nematic phase does not arise.  Since the nematic phase has a lower density 
  of states, it is expected that in mean-field theory, it will also have a lower entropy at temperatures small compared to $\mu$.   }
   \label{fig3}
\end{figure} 

\begin{figure}
\includegraphics[width=6.cm]{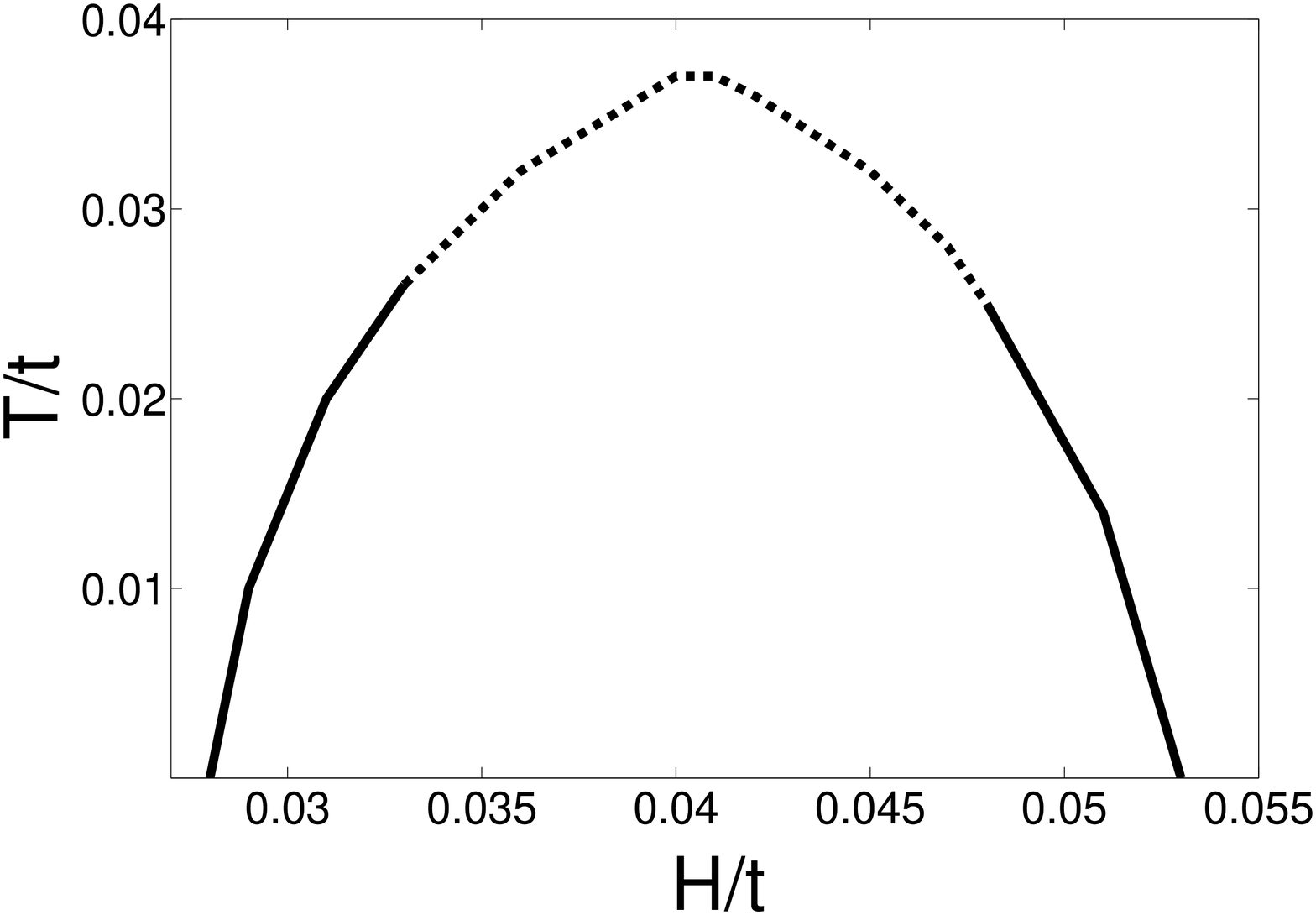}
\caption{Phase diagram of the system  without spin-orbit coupling showing the finite temperature 
boundaries of the nematic phase.  First-order transitions (solid curve) at low temperatures give way 
to continuous transitions (dashed curve) at higher temperatures as the vH singularities get smoothed out.  Finite temperature metamagnetic cross-overs away from the nematic phase are not shown.  Here, $U=V=2.5t$, $\mu = 0.81t$ and $\lambda = 0.2t$.    }
\label{pdHT}
\end{figure}

In Fig. \ref{fig3}, the single quasi-particle density of states is plotted in the nematic phase.  There is a 
double peak near the Fermi level, whose splitting corresponds to the size of the nematic gap.  
Also shown in Fig. \ref{fig3} is the density of states at a slightly lower field, where the system is just 
about to enter the nematic phase.  We note that there is a reduction of the quasiparticle density 
of states at the Fermi level in the nematic phase; this in turn implies that the entropy of the nematic phase at low temperatures 
within mean-field theory is lower than the neighboring isotropic phases.  This is also seen 
in Fig. \ref{pdHT} from the curvature of the phase boundaries of the nematic phase at finite temperatures.  
We highlight here that this result from mean-field theory disagrees with the experimental observations 
reported  in Ref.  \cite{Borzi2007}, where the nematic phase boundaries ``fan" outward, implying  that in actuality, 
the nematic phase has a higher entropy than the neighboring isotropic phases.  We believe 
that this effect must stem from fluctuations (both thermal and quantum) due to the presence of 
nematic domains.

\section{Effect of spin-orbit coupling: angle-dependent metanematic transitions}
\label{soc}
The results above were based on the assumption that spin $SU(2)$ symmetry is preserved in the system.  
However, the experimental phase diagram \cite{Grigera2003} of \SRO (Fig. \ref{expt}) exhibits an $O(1)$ anisotropy 
in the critical field at which a metamagnetism occurs: the critical field with H along the $c$ axis is about $1.6$ times 
larger than the case when it points in the $ab$ plane.  A natural way to account for this anisotropy is 
to consider the effect of spin-orbit coupling in the material.  Since the material has an inversion center, 
k-dependent spin-orbit interactions, such as the Rashba coupling, are forbidden, and we proceed by 
considering the effect of an atomic  spin-orbit coupling, given in equation \ref{hso}.  
While the precise magnitude of the spin-orbit coupling constant 
$\lambda_{s.o.}$ for this material is unclear, LDA calculations of the monoloyer ruthenate 
compound 
${\rm Sr}_2{\rm RuO}_4$, suggest that  $\lambda_{s.o.}  $ is approximately $10$ percent of $E_f$
\cite{Haverkort2008, Liu2008}.  Indeed, in recent angle-resolved photoemission (ARPES) studies of \SRO, 
LDA calculations had to employ spin-orbit coupling in order to fit properly to the Fermi surface \cite{Tamai2008}.  Since this spin-orbit coupling term is onsite, 
it seems reasonable to expect that similar values 
also hold for the bilayer compound.   

When such spin-orbit coupling terms are included in the Hamiltonian, 
the $d_{xy}$ 
band 
must necessarily be taken 
into account.  However, since the spin-orbit coupling is smaller than the crystal field splitting, we can 
treat it as a perturbation, and project the angular momentum operator $\bm L$ 
onto the $t_{2g}$ subspace.  
To be more explicit, a valid choice of the orbital 
angular momentum operators projected onto the $t_{2g}$ manifold is

\begin{eqnarray}
\ell^x = \left( \begin{array}{ccc}
0 & 0 & 0 \\
0 & 0 & -1 \\
0 & -1 & 0 \end{array} \right), && \ 
\ell^y = \left( \begin{array}{ccc}
0 & 0 & -1 \\
0 & 0 & 0 \\
-1 & 0 & 0 \end{array} \right), \nonumber \\
 \ell^z &=& -i \left[ \ell^x, \ell^y \right] . 
\end{eqnarray}
In terms of these operators, the spin-orbit coupling in each layer of our system is represented by the $6 \times 6$ matrix:
\begin{equation}
\bm L \cdot \bm S = \left( \begin{array}{cc}
 \ell^z & \ell^x - i \ell^y \\
 \ell^x + i \ell^y  & - \ell^z \end{array}\right) . 
 \end{equation}

\begin{figure}
\includegraphics[width=8.cm]{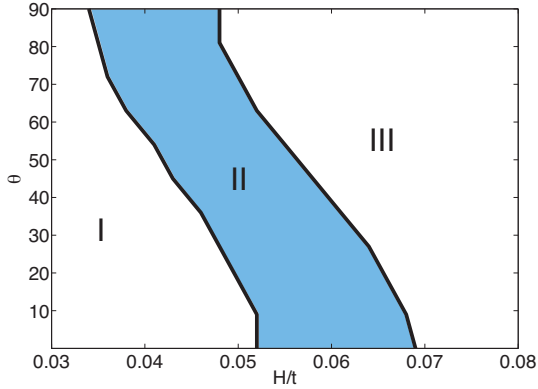}
\caption{ (Color online) Angle-dependent metanematic phase boundaries  
in the presence of spin-orbit coupling showing an $O(1)$ anisotropy in the critical fields.  
When the field is tilted towards the {\it ab} plane, the $C_4$ point group symmetry is 
automatically broken.  In region I at low fields, the moment is small and there is also a small 
nematic phase present for all $\theta > 0$.  The first metamagnetic transition into region II gives rise to an 
accompanying jump in nematicity in addition to the magnetization.  Finally, the second metamagnetic 
transition into region III results in a discontinuous decrease in the size of the nematic order parameter 
as well as a jump in magnetization.  
Here, $U=V=2.5t$, $\mu = 0.81t$ and $\lambda_{s.o.} = 0.2t$.    }
   \label{angle}
\end{figure}

Having in mind the framework of a minimal model which captures the essential features 
of the experimental phase diagram, we first treat the quasi-2D band as a free electron system and neglect 
the effect of Coulomb interactions.  We have checked that the inclusion of a Hubbard repulsion on this 
band produces no qualitative changes to the results reported here.  Furthermore, we neglect the terms in $H_2$ as before.  
This way, the mean-field order parameters in the presence of spin-orbit coupling are identical 
to those in Eq. \ref{op}, and the spin quantization axis is defined to be in the direction of the applied field.
However, the spin-orbit coupling term is sensitive to the orientation of the magnetic field relative to the c-axis since 
the orbital angular momentum operators above are defined with respect to the crystalline axis of the system.  
More explicitly, in the tilted field, the spin-orbit term is modified as 
\begin{eqnarray}
H_{so}(\theta) &=& \lambda_{s.o.} \bm L \cdot \tilde{\bm S}(\theta) \nonumber \\ 
\tilde{ \bm S}(\theta) &=& \exp{ \left(-i \frac{\vec{\sigma} \cdot \vec{n} }{2} \theta \right) } \bm S 
\exp{ \left(i \frac{\vec{\sigma} \cdot \vec{n} }{2} \theta \right) },
\end{eqnarray}
where $\theta $ is the angle of the applied field relative to the c-axis and $\vec{n}$ is a
 unit vector either in the crystalline {\it a} or {\it b} direction.

In Fig. \ref{angle}, we show the 
phase diagram in the $H-\theta$ plane, keeping $U=V$ fixed and $\lambda_{s.o.} = 0.2t$.    As 
the field is tilted towards the ab plane, we see that the critical field at which the metamagnetic transition 
occurs is smaller in magnitude.  For all angles, there are two metamagnetic transitions with a nematic 
phase in between.  When the field is tilted away from the c-axis, the crystalline $C_4$ symmetry is 
broken and we distinguish a phase in which the nematic order jumps from a small value (region I) 
to a large value (region II) and jumps back to a small value (region III).  Thus the tilted field 
shows the nematic analog of metamagnetic transitions, and we refer to these transitions here as 
``metanematic" transitions \cite{Puetter2007}.  We note that although we find metamagnetic transitions with the 
same general angle-dependence as seen in experiments, the nematic phase in our model 
does not have the correct topology in the field-angle plane (compare Fig. \ref{angle} and Fig. \ref{expt}).  
Instead, we find that if a nematic phase occurs, it remains present for 
all field orientations.  
Given the structure of the phases shown in Fig. \ref{phasediagram}b, it does not appear 
impossible that the correct topology could emerge under appropriate circumstances, but in the present 
model, this would require
angle-dependent interactions.  While such terms are unphysical at the bare microscopic level, they could arise at the effective level due to fluctuation effects 
that have not been considered here.

The ordering tendencies which we have considered so far do not break lattice translational symmetry.  
However, inelastic neutron scattering experiments \cite{Capogna2003, Ramos2008} have 
provided evidence that there are substantial incommensurate spin fluctuations in this system, 
although static spin-density-wave (SDW) order has not been observed at zero 
external field.  The incommensurate spin fluctuations in this system occur 
primarily due to the partial nesting of the Fermi surface.  Therefore, we must check whether 
such finite $\bm q$ ordering tendencies in this material are favored over the uniform nematic order 
proposed in this paper.  To do this, we have computed  the generalized one-loop spin susceptibility
\begin{eqnarray}
\left[ \chi^{i j}(\bm q) \right]^{s t}_{b a} && = \int_0^{\beta} d \tau \sum_{\bm p \bm p'} \sum_{\alpha \beta \gamma \delta} \nonumber \sigma^i_{\alpha \beta} \sigma^j_{\gamma \delta}  \times \\
&& \langle T_{\tau} d^{\dagger}_{s \bm p \alpha}(\tau) d_{t \bm p + \bm q \beta } (\tau) d^{\dagger}_{a \bm p' \gamma} (0) d_{b \bm p' - \bm q  \delta}(0) \rangle  \nonumber \\
\end{eqnarray}
in the presence of spin-orbit coupling and magnetic field. The electron Hamiltonian is a $ 12 \times 12$ matrix, and the electron propagators are spin and orbital
dependent 
which in turn makes the above spin susceptibility a $ 36 \times 36 $ matrix.   
Using the random phase approximation (RPA), we find that an instability towards the formation of SDW 
order occurs in the system (taking $U=V$) when U exceeds a critical strength $U_{c,sdw}$ which satisfies 
\begin{equation}
\frac{U_{c,sdw}}{2} {\rm Max}  \left( {\rm eig} \left[  \chi \right]   \right)  = 1
\end{equation}
and for the band-structure parameters we have been using ($\lambda_{s.o.} = 0.2 t$, $t'=0$, $\mu = 0.81t$) the maximal eigenvalue of the susceptibility matrix is obtained at the wave-vector $ \bm q = \left( 0.27, 0.27 \right) \pi $ where the spin susceptibility $\chi^{zz}(\bm q)$ obtains its largest value.  From this mean-field estimation, we find that $U_{c,sdw} \approx 3.1t$, which is also close to the critical coupling required for Ferromagnetism, as seen in Figure 
\ref{phasediagram}.  

Thus, in addition to accounting for the angle-dependent metamagnetic transitions observed in this system, the inclusion of spin-orbit coupling also naturally explains why static incommensurate 
SDW order does not occur despite the presence of incommensurate spin fluctuations  in this system.

\section{Discussion}

We have shown in this paper that the remarkable low temperature properties of \SRO 
can be understood as being a consequence of orbital ordering of the quasi-1D bands.  
We have been able to account for most of the gross features of the experiments 
from a simple microscopic model.  
Our results are based on a mean-field solution of a system with strong electron interactions and a justification for focusing solely on this approach has not been provided.  Although 
metamagnetism and nematicity generally arise as strong coupling effects in metals and are therefore difficult to treat in a controlled theoretical fashion, 
 the divergence of the density of states $\nu$ at the  
vH point 
insures that the Stoner criterion can be satisfied even for  weak interactions, which are  much less than the bandwidth.  Thus, by the notion of 
adiabatic continuity, it may be legitimate to treat this problem from a weak-coupling standpoint.      

Previous attempts to explain metamagnetism and nematicity in this system \cite{Kee2005, Honerkamp2005, Berridge2009} were based on the
assumption that the quasi-2D bands drive the nematic and metamagnetic transitions.  By contrast, 
we point out here that it is much more natural to think instead that the quasi-1D bands are 
responsible for the transitions.  We note, from a symmetry standpoint, that the quasi-1D bands form a 
two-fold representation of the $C_4$ rotation symmetry of this system: thus, the nematic phase which 
breaks $C_4$ rotation symmetry corresponds to an orbital ordering among these bands.  Furthermore, 
it is known from experiment that the monolayer ruthenate ${\rm Sr}_2{\rm Ru}{\rm O}_4$ does not exhibit 
metamagnetism for magnetic fields upto 30 Tesla.  The primary difference in the electronic structure 
of the monolayer vs. the bilayer compound is the large bilayer splitting in the latter.  We have shown 
here that while the quasi-2D bands are only weakly affected by the bilayer splitting, the quasi-1D bands 
are rather strongly affected by it.  Thus, the experimental differences between the monolayer and bilayer 
ruthenate compounds as well as symmetry considerations lead us to propose that quasi-1D bands in this system 
are primarily responsible for the rich phase diagram of the bilayer ruthenate.  
Although we propose a different microscopic origin for the Fermi surface distortion from that used in previous work on the subject, we note that our picture is still one of a weak-coupling Pomeranchuk type.  As such it automatically retains the attractive feature, common to any weak anisotropic distortion of electronic structure in the vicinity of $E_F$, that it naturally predicts a sensitivity to impurity scattering \cite{Ho2008}.  
This in turn matches one of the key experimentally determined characteristics of the behavior that we set out to explain.

Along with the focus on the 1D bands comes the existence of a nesting vector, $2k_F$, 
and the issue of SDW order.  At mean-field level, the preferred SDW instability for a pair of 
orthogonal quasi-1D bands leads \cite{Yao2006} either to bidirectional order with ordering vectors 
$(2k_F, \pi)$ and $(\pi, 2k_F)$, or unidirectional order with ordering vector $(2k_F, 2k_F)$.  
However, inelastic neutron scattering studies have found that the most dominant peaks are observed 
in the $(2 k_F, 0)$ direction \cite{Capogna2003}.  
Nevertheless, we have shown that even moderate spin-orbit coupling reduces the nesting to the extent that the {\it mean field} tendency to nematic 
and SDW ordering are comparably strong.  Moreover, we expect fluctuation effects to reduce the 
ordering tendency of an incommensurate SDW (which has a gapless sliding mode and two 
nearly gapless spin-wave modes) relative to the Ising-like nematic order, especially 
given the quasi-2D structure of this material, which makes fluctuational corrections all the more 
significant.

However, there are still two qualitatively important features of the experimental data that are not well 
accounted for by any mean-field treatment we know of:  1)  The intermediate nematic phase has been 
shown, in experiment, to have higher entropy at low temperatures than either of the adjacent  disordered 
phases.  In contrast, at mean-field level, the ordered phase always has a lower Fermi surface density of 
states than the proximate disordered phases, and so has smaller low temperature entropy.  2)  
Doping studies suggest that at least part of the peaks in the density of states that are thought to drive the metamagnetism are somehow locked to the Fermi energy, rather than being purely features of an underlying rigid band-structure.
These features we believe are signatures of strong-coupling effects, and can be 
accounted for by a more sophisticated theory in which local nematic order is present over a broad range 
of magnetic fields (including $B=0$), but only propagates to long distances in the narrow range of $B$ 
in which macroscopic anisotropies are observed.  

More generally, it is clear that fluctuation effects play a significant role in the physics.  Even though the 
putative metamagnetic quantum critical point is preempted by the nematic phase, the 
observed fluctuational phenomena that lead to the conjectured critical point in the first place remain 
real and dramatic.  It is likely that they reflect the existence of a ``nearby" quantum critical point, 
even if it is not actually observed.  In this context, mean-field results, of the sort discussed in the present 
paper, should be adopted with caution.  At the very least, the effective parameters that enter 
our model must be reinterpreted as strongly renormalized effective parameters, given that the 
observed bandwidths  \cite{Reference} are narrower by a factor of order 10-100 than the bandwidths 
found in LDA calculations \cite{Singh2001}.  We defer the fascinating study of fluctuation effects 
to a future publication.    

Finally, we make a comment about the observed resistivity anisotropy in this material when $H \parallel c$.  While we have presented a symmetry argument for why such a resistivity anisotropy ought to be 
present, we have not explicitly computed the resistivity anisotropy from our model.  
Indeed, any small nematic Fermi surface distortion like that discussed here is unlikely to produce a large effect on the low-temperature resistivity due to the distortion alone. 
The reason for this, is that the nematic phase arises from a discrepancy in 
electrons which are close to a vH singularity; these, in turn, have a very small 
characteristic velocity and hence would contribute most weakly to transport 
signatures.    The distortion of the Fermi surface is more likely to change the transport properties by being a source of domains and domain-wall scattering, something which is beyond the scope of the calculations that we report.
We note further that in the experiments of Ref. \cite{Borzi2007}, 
when the field is tilted towards the {\it ab} plane, both the average resistivity and the anisotropy rapidly 
decrease with angle.  In our picture, we imagine that as the field is tilted, nematic domains get aligned 
and scattering is therefore considerably reduced.   Thus, ironically, precisely the same signal which 
was used to detect the nematic fluid would prove to be useless deep in the nematic phase, when 
the system forms a single macroscopic nematic order.  A more quantitative theory of the physics 
discussed here will be presented elsewhere.  

\begin{acknowledgments}
We are grateful to M. Allan, J. C. Davis, E. Fradkin,  H.-Y. Kee, R. B. Laughlin,  M. Lawler,  Y. Maeno, V. Oganesyan, A. Rost, D. J. Scalapino,  K. Shen, D. Singh, and H. Yao for insightful discussions.  
This work was supported in part by the Cornell Center for Materials Research through NSF Grant No. 0520404 (EAK), the Sloan Foundation, NSERC of Canada and the Ontario ERA (AP), NSF DMR  0758356 (SAK) and the Stanford Institute for Theoretical Physics (SR).   

\end{acknowledgments} 

{\it Note added - } As we were preparing to submit this paper, we became aware of a related study by 
W-.C. Lee and C. Wu, Ref. \cite{Lee2009}.  

\appendix*
\section{Metamagnetism and nematicity of quasi-1D bands}
Here, we present, for the sake of clarity, the derivation of the Mean-field equations which were used to deduce the phase diagram in Fig. \ref{phasediagram}.  The mean-field order parameters 
are defined via 
\begin{eqnarray}
\langle n_{x, \lambda, {\vec R},  \sigma } \rangle &=& \frac{1}{4} \left( N^{\lambda} + N_{o}^{\lambda} + \sigma M^{\lambda} + \sigma N_s^{ \lambda} \right) \nonumber \\
\langle n_{y , \lambda. {\vec R},  \sigma } \rangle &=& \frac{1}{4} \left( N^{\lambda} - N_{o}^{\lambda} +  \sigma M^{\lambda} - \sigma N_s^{ \lambda} \right) 
\end{eqnarray}
After decomposing the interactions in terms of the above expectation values, we 
arrive at the following one-particle Hamiltonian:
\begin{eqnarray}
&& H(\vec{k}) = \left(\begin{array}{cc}
H_{x} (\vec{k})& 0 \\
0 & H_{y} (\vec{k})\end{array} \right) 
\end{eqnarray} 
where
\begin{eqnarray}
&& H_{a}(\vec{k}) =-2t \cos{k_a} \hat{1}_{4 \times 4}  \nonumber + \\ && \left( \begin{array}{cccc}
\delta \mu_{a, \uparrow}^{(1)} - h& 0 & -t & 0  \\
0 & \delta \mu_{a, \downarrow}^{(1)}+h& 0 & -t  \\
-t & 0 & \delta \mu_{a, \uparrow}^{(-1)}-h& 0  \\
0 & -t & 0 &\delta \mu_{a, \downarrow}^{(-1)} +h \end{array} \right) \nonumber 
\end{eqnarray}
and $a = x,y$.
We have also defined the quantities
\begin{eqnarray}
\delta \mu_{x, \sigma}^{(\lambda)} &=& \frac{U}{4} \left( N_o^{\lambda} - \sigma M^{\lambda} - \sigma N_s^{\lambda} \right)
 - \frac{V}{2}N_o^{\lambda}    \nonumber \\
\delta \mu_{y ,\sigma}^{(\lambda)} &=&  \frac{U}{4} \left( -N_o^{\lambda} - \sigma M^{\lambda} + \sigma N_s^{\lambda} \right)
 + \frac{V}{2}N_o^{\lambda}  
\end{eqnarray}
Thus, by neglecting the hybridization between the $x,y$ orbitals, the Mean-field Hamiltonian takes 
the Block-diagonal form above and the quasiparticle bonding and anti-bonding energies are easily obtained:
\begin{eqnarray}
   \epsilon^{x}_{\pm \sigma} &=& -2t \cos k_x - \sigma h+ \frac{1}{2} \left(  \delta \mu^{(1)}_{x ,\sigma} + \delta \mu^{(-1)}_{x, \sigma} \right) \nonumber \\ && \pm \sqrt{t^2+ \frac{1}{4} \left( \delta \mu^{(1)}_{x ,\sigma} - \delta \mu^{(-1)}_{x, \sigma} \right)^2 } \nonumber \\
 \epsilon^{y}_{\pm \sigma} &=& -2t \cos k_y - \sigma h+ \frac{1}{2} \left(  \delta \mu^{(1)}_{y ,\sigma} + \delta \mu^{(-1)}_{y, \sigma} \right)  \nonumber \\ && \pm \sqrt{t^2+ \frac{1}{4} \left( \delta \mu^{(1)}_{y ,\sigma} - \delta \mu^{(-1)}_{y, \sigma} \right)^2 } 
\end{eqnarray}
Due to the 1D band dispersion, it is possible to obtain the density of states and the grand-canonical free energy density 
analytically by  summing over the 8 quasi-particle bands of this model.  At zero temperature,
\begin{eqnarray}
&&F = F_0(M^{\lambda},N^{\lambda}_s,N^{\lambda}_o) + \int_{-\infty}^{\mu} \left( E- \mu \right) \nu(E) d E  \nonumber \\
&&F_0(M^{\lambda},N^{\lambda}_s,N^{\lambda}_o)= \frac{U}{8} \sum_{\lambda} \left[ \left( \bar{N}^{\lambda  } \right)^2 + \left( M^{\lambda } \right)^2 + \left(N_s^{\lambda } \right)^2 -\left( N_o^{\lambda } \right)^2 \right] \nonumber \\
&& + \frac{V}{4} \sum_{\lambda} \left[ \left( \bar{N}^{\lambda } \right)^2 + \left( N_o^{\lambda } \right)^2 \right]  \nonumber \\
\end{eqnarray}
\begin{widetext}
\begin{eqnarray}
&& \int_{-\infty}^{\mu} \left( E- \mu \right) \nu(E) d E  = - \frac{2}{\pi} {\rm Re} \sum_{ a=x,y} \sum_{ \sigma, \sigma' = \pm 1}   \left[ 1 - \frac{1}{4} \left(
 \mu + \sigma h-\frac{1}{2}\left( \delta \mu_{a, \sigma}^{(1)} + \delta \mu_{a, \sigma}^{(-1)} \right) + \sigma' \sqrt{t^2 +\frac{1}{4}  \left(\delta \mu_{a, \sigma}^{(1)} - \delta \mu_{a, \sigma}^{(-1)} \right)^2} \right)^2  \right]^{1/2}  \nonumber \\
&& - \frac{2}{\pi} {\rm Re} \sum_{\sigma, a=x,y} \sum_{ \sigma'= \pm 1}  \frac{1}{2}\left(
 \mu + \sigma h-\frac{1}{2}\left( \delta \mu_{a, \sigma}^{(1)} + \delta \mu_{a, \sigma}^{(-1)} \right) + \sigma' \sqrt{t^2 +\frac{1}{4}  \left(\delta \mu_{a, \sigma}^{(1)} - \delta \mu_{a, \sigma}^{(-1)} \right)^2} \right)^2   \times\nonumber \\
 &&  \sin^{-1} \left[ \frac{\mu + \sigma h-\frac{1}{2}\left( \delta \mu_{a, \sigma}^{(1)} + \delta \mu_{a, \sigma}^{(-1)} \right) + \sigma' \sqrt{t^2 +\frac{1}{4}  \left(\delta \mu_{a, \sigma}^{(1)} - \delta \mu_{a, \sigma}^{(-1)} \right)^{2}}}{2} \right]  \nonumber \\
\end{eqnarray}
\end{widetext}

The phase diagram is then obtained by minimizing the above free energy with respect to the 
order parameters.  We note that even upon including curvature effects in the band-structure, the grand canonical free energy density can be expressed analytically in terms of complete Elliptic Integrals.  However, for the sake 
of simplicity, we don't include such terms here.  While such Fermi terms do modify the precise 
location of phase boundaries, they have no qualitative effect on the physics.

\bibliography{SrRuO327}
\end{document}